\documentclass[aps,twocolumn,pra,showpacs,superscriptaddress,tightenlines]{revtex4-2}
\usepackage{amsmath}
\usepackage{graphicx}
\usepackage{color}
\usepackage{amsfonts}
\usepackage[colorlinks,citecolor=blue]{hyperref}
\begin{document}

\title{Optimized mechanical quadrature squeezing beyond the 3-dB limit via a gradient-descent algorithm}
\author{Yu-Hong Liu}
\affiliation{Key Laboratory of Low-Dimensional Quantum Structures and Quantum Control of
Ministry of Education, Key Laboratory for Matter Microstructure and Function of Hunan Province, Department of Physics and Synergetic Innovation Center for Quantum Effects and Applications, Hunan Normal University, Changsha, 410081, China}

\author{Jie-Qiao Liao}
\email{Corresponding author: jqliao@hunnu.edu.cn}
\affiliation{Key Laboratory of Low-Dimensional Quantum Structures and Quantum Control of
Ministry of Education, Key Laboratory for Matter Microstructure and Function of Hunan Province, Department of Physics and Synergetic Innovation Center for Quantum Effects and Applications, Hunan Normal University, Changsha, 410081, China}
\affiliation{Institute of Interdisciplinary Studies, Hunan Normal University, Changsha, 410081, China}

\begin{abstract}
The preparation of mechanical quadrature-squeezed states holds significant importance in cavity optomechanics because the squeezed states have extensive applications in understanding fundamental quantum mechanics and exploiting modern quantum technology. Here, we propose a reliable scheme for generating mechanical quadrature squeezing in a typical cavity optomechanical system via seeking optimal cavity-field driving pulses using the gradient-descent algorithm. We realize strong quadrature squeezing in a mechanical resonator that exceeds the $3$-dB steady-state limit, even with a thermal phonon occupancy of 100. Furthermore, the mechanical squeezing can be ultrarapidly created within one mechanical oscillation period. We also obtain the optimal pulsed drivings associated with the created mechanical squeezings and analyze the mechanism for mechanical squeezing generation. This paper will promote the application of optimal quantum control in quantum optics and quantum information science.
\end{abstract}
\date{\today}
\maketitle

\section{INTRODUCTION}
Cavity optomechanics~\cite{Kippenberg2014RMP}
is a frontier research field focusing on the radiation-pressure interaction
between cavity fields and moving mirrors at macroscopic scales. Owing to the
nonlinear properties of the optomechanical interaction and the development
of laser cooling techniques~\cite%
{Wilson-Rae2007PRL,Marquardt2007PRL,ChanNature2011,TeufelNature2011},
optomechanical systems provide new paths and opportunities for generation
and manipulation of macroscopic nonclassical states, such as entangled
states~\cite%
{VitaliPRL2007,Palomaki2013Science,Riedinger2016Nature,Jiao2020PRL,Yu2020Nature,Lai2022PRL}%
, quantum superposed states~\cite%
{Bose1997PRA,Marshall2003PRL,LiaoPRL2016,LiaoPRA2020}, and squeezed
states of fields~\cite{
FabrePRA1994,ManciniPRA1994,Brooks2012,PurdyPRx2013,Safavi-Naeini2013,AgarwalNJP2014}
and mechanical resonators~\cite%
{WoolleyPRA2008,HuangPRA2023,TeufelNN2009,NunnenkampPRA2010,LiaoPRA2011,MilburnOL1981,MariPRL2009,PontinPRL2014,PontinPRL2016,ChowdhuryPRL2020,YongChun2023PRA,ClerkNJP2008,SzorkovszkyPRL2011,VasilakisNP2015,MengPRL2020,PirkkalainenPRL2015,GenesPRA2009,AgarwalPRA2016,AsjadPRA2014,LuPRA2015,RashidPRL2016,KronwaldPRA2013,XiongOL2018}.

Quantum squeezing, as an essential quantum resource, plays an increasingly
significant role in modern quantum technology ranging from quantum precision measurement to quantum information processing~\cite%
{Braunstein2003,Drummond2004}. Quantum
squeezing of mechanical modes holds significance as it has the potential to
enhance the accuracy of quantum measurements~\cite{CavesRMP1980}. There
exist many theoretical and experimental schemes for generating mechanical
squeezing based on various methods, such as parametric squeezing~\cite%
{NunnenkampPRA2010,LiaoPRA2011,
MilburnOL1981,MariPRL2009,PontinPRL2014,PontinPRL2016,ChowdhuryPRL2020,YongChun2023PRA}%
, quantum measurement~\cite%
{ClerkNJP2008,SzorkovszkyPRL2011,VasilakisNP2015,MengPRL2020,PirkkalainenPRL2015}%
, quantum state transfer~\cite{GenesPRA2009,AgarwalPRA2016,HuangPRA2023}, mechanical
nonlinearities~\cite{NunnenkampPRA2010,AsjadPRA2014,LuPRA2015},
and reservoir engineering~\cite{KronwaldPRA2013,XiongOL2018}. In particular, some methods~\cite{LiaoPRA2011,PontinPRL2014,PontinPRL2016,SzorkovszkyPRL2011,GenesPRA2009,AsjadPRA2014,LuPRA2015,KronwaldPRA2013} can be used to generate the quadrature squeezing, which exceeds the 3-dB squeezing limit~\cite{MilburnOL1981,Scully1997}. The strong squeezing has wide applications in quantum technology, and its generation is a significant task in quantum optics. In addition, from the view
point of realistic application, an ultrafast generation of strong mechanical squeezing remains a challenge.

In cavity optomechanical systems, the external optical driving provides an
effective way to control the quantum properties and dynamical behaviors of
the system~\cite%
{VannerPNAS2011,VannerNC2013,LiaoPRA2011,MariPRL2009,MeenehanPRX2015}%
. The design of an optimal optical driving to achieve a supposed goal is an interesting topic in this system. Currently, quantum control
techniques have been successfully applied to various schemes in cavity optomechanical systems~\cite%
{Hammerer2017AIMO}%
, such as robust state transfer~\cite{Wang2021AP}, optimized cooling~\cite{Jacobs2011,MagriniNature2021,LiuPRA2022}, entanglement enhancement~\cite%
{Stefanatos2017QST,ClarkeNJP2020}, and strong squeezing~\cite{Xiong2020PR,Basilewitsch2019AQE}. Furthermore, optimal control technology based on the gradient-descent algorithm~\cite{Dong2020} has been
recognized as a powerful method to accomplish complex control tasks, and hence it
is expected to provide new ways to optimize the control of cavity
optomechanical systems.

In this paper, we apply the  gradient-descent algorithm to prepare mechanical quadrature squeezing in a typical cavity optomechanical system. Our
scheme can break the $3$-dB steady-state squeezing limit even when the environment thermal phonon
occupation associated with the mechanical resonator reaches the order of 100. The mechanical quadrature squeezing is achieved by designing a proper pulsed driving to the cavity field.
Concretely, we use the gradient-descent algorithm to iteratively optimize
the pulsed driving for achieving strong mechanical squeezing. The optimal waveforms
of the pulsed driving amplitude and phase are also obtained. In particular,
mechanical squeezing can be ultrarapidly prepared within one mechanical oscillation
period, providing more chance for realistic applications of the generated squeezing before decoherence. Our method will encourage further researches on optimal quantum control in cavity optomechanical systems.

The rest of this paper is organized as follows. In Sec.~\ref{12}, we
introduce the Hamiltonians and present the equations of motion. In Sec.~\ref%
{13}, we show the generation of mechanical quadrature squeezing using the gradient-descent algorithm. In Sec.~\ref{14}, we present some discussions concerning the influences of driving amplitude and phase noises on the squeezing generation, the physical mechanism of the present squeezing-generation method, the experimental implementation of this squeezing-generation scheme, and the applications of the optimization methods in cavity optomechanics. Finally, we conclude this paper in Sec.~\ref{15}. The Appendix shows the
derivation of the variation $\delta [\Delta X_{b}^{2}(\theta ,T)]/\delta \mathcal{Q}_{m}$ used in the gradient-descent algorithm.

\section{HAMILTONIANS AND EQUATIONS OF MOTION\label{12}}
We consider a typical cavity optomechanical system that consists of a
mechanical resonator optomechanically coupled to a single-mode cavity field.
To control the dynamics of the system, the cavity mode is driven by a strong
pulsed field with carrier frequency $\omega _{\mathrm{L}}$, driving
amplitude $\Omega (t)$, and driving phase $\phi (t)$. In a rotating frame
defined by the unitary operator $\exp (-i\omega _{\mathrm{L}}ta^{\dag }a)$,
the Hamiltonian of the system reads ($\hbar =1$)
\begin{align}\label{eq3}
H_{\mathrm{opt}}(t) &=\Delta _{\mathrm{c}}a^{\dag }a+\omega _{\mathrm{m}%
}b^{\dag }b-g_{0}a^{\dag }a(b^{\dag }+b) \notag \\ &\hspace{0.3cm}+[\Omega (t)\mathrm{e}^{-i\phi (t)}a^{\dag }+\mathrm{H.c.}],
\end{align}%
where $a^{\dag} $ $(a)$ and $b^{\dag }$ $(b)$ are, respectively, the
creation (annihilation) operators of the cavity field and the mechanical
resonator, with the corresponding resonance frequencies $\omega _{\mathrm{c}}$
and $\omega _{\mathrm{m}}$. The $\Delta _{\mathrm{c}}=\omega _{\mathrm{c}%
}-\omega _{\mathrm{L}}$ is the detuning of the cavity-field resonance
frequency with respect to the pulsed-field carrier frequency. The $g_{0}$
term describes the radiation-pressure coupling between the cavity field and the mechanical resonator, with $g_{0}$ being the single-photon optomechanical-coupling strength.

In the open-system case, we assume that the cavity field is coupled to a
vacuum bath, while the mechanical resonator is connected to a heat bath.
Considering the Markovian dissipations, the evolution of the system is
governed by the quantum master equation
\begin{equation}
\dot{\rho}\!=\!i[\rho ,H_{\mathrm{opt}}(t)]\!+\kappa \mathcal{D}[a]\rho
+\!\gamma (\bar{n}_{\mathrm{m}}+1)\mathcal{D}[b]\rho +\!\gamma \bar{n}_{%
\mathrm{m}}\mathcal{D}[b^{\dag }]\rho ,  \label{Eq2}
\end{equation}%
where $\rho $ is the density matrix of the optomechanical system, $\mathcal{D%
}[o]\rho =o\rho o^{\dag }-(o^{\dag }o\rho +\rho o^{\dag }o)/2$ (for $o=a$, $%
a^{\dag }$, $b$, and $b^{\dag }$) is the Lindblad superoperator~\cite%
{Agarwal2013}, and $H_{\mathrm{opt}}(t)$ is defined in Eq.~(\ref{eq3}).~The
parameters $\kappa $ and $\gamma $ are, respectively, the damping rates of
the cavity field and the mechanical resonator, and $\bar{n}_{\mathrm{m}}$
is the environmental thermal-excitation occupation of the mechanical
resonator.

Considering the strong-driving case of the optomechanical cavity, then the dynamics of the system can be linearized. To this end, we adopt the displacement-transformation method to
separate the semiclassical motion and quantum fluctuation. Concretely,
we perform the displacement transformations $D_{a}(\alpha )=\exp (\alpha
a^{\dagger}-\alpha ^{\ast }a)$ and $D_{b}(\beta )=\exp (\beta b^{\dagger
}-\beta^{\ast }b)$ for the density operator $\rho(t)$, namely,
\begin{equation}
\rho^{\prime }(t)=D_{a}(\alpha )D_{b}(\beta )\rho(t) D_{b}^{\dagger }(\beta
)D_{a}^{\dagger }(\alpha ),  \label{s1eq2}
\end{equation}
where $\rho^{\prime }(t)$ represents the density operator in the displaced representation, and $\alpha(t)$ and $\beta(t)$ are the displacement amplitudes of the cavity field and
the mechanical resonator, respectively. By substituting $\rho(t)=D_{b}^{\dagger }(\beta
)D_{a}^{\dagger }(\alpha )\rho^{\prime }(t)D_{a}(\alpha )D_{b}(\beta )$ into Eq.~(\ref{Eq2}) and setting the
coefficients of the driving terms to be zero, we obtain the quantum master
equation in the displaced representation as
\begin{equation}  \label{Eq45}
\dot{\rho}^{\prime } =i[\rho^{\prime },H_{\mathrm{dis}}(t)]+\kappa \mathcal{D%
}[a]\rho^{\prime }+\gamma ( \bar{n}_{\mathrm{m}}+1) \mathcal{D}[ b]
\rho^{\prime }+\gamma\bar{n}_{\mathrm{m}}\mathcal{D}[b^{\dag }] \rho^{\prime
}, \\
\end{equation}%
where $H_{\mathrm{dis}}(t)$  is the Hamiltonian in the displaced representation, defined as
\begin{align}
H_{\mathrm{dis}}(t)&=\Delta (t)a^{\dagger }a+\omega _{\mathrm{m}%
}b^{\dagger }b -g_{0}a^{\dagger }a(b+b^{\dagger })\notag \\ &\hspace{0.3cm}+[G(t)a^{\dagger}+G^{\ast }(t)a](b+b^{\dagger }) .  \label{eq5}
\end{align}%
In Eq.~(\ref{eq5}), we introduce the linearized optomechanical-coupling strength $%
G(t)=g_{0}\alpha(t)$ and the normalized driving detuning $%
\Delta(t)=\Delta_{c}+g_{0}[\beta(t)+\beta^{*}(t)]$. The two
displacement amplitudes $\alpha(t)$ and $\beta(t)$ are governed by the
equations of motion,
\begin{subequations}
\begin{align}
\dot{\alpha}(t)&=-\left[i\Delta(t)+\frac{\kappa}{2}\right]\alpha(t)+i\Omega(t)\text{e}%
^{-i\phi(t)} ,  \label{eq2a} \\
\dot{\beta}(t)&=-\left(i\omega_{\mathrm{m}}+\frac{\gamma}{2}\right)\beta(t)-ig_{0}|%
\alpha(t)|^{2} .  \label{eq2b}
\end{align}
\end{subequations}
In the strong-driving case, $|\alpha(t)| \gg 1$, we can safely omit the $g_{0}$
term in Eq.~(\ref{eq5}), and obtain the linearized optomechanical
Hamiltonian
\begin{equation}
H_{\mathrm{lin}}(t)=\Delta (t)a^{\dagger }a+\omega _{\mathrm{m}}b^{\dagger
}b +[G(t)a^{\dagger}+G^{\ast }(t)a](b+b^{\dagger }) ,  \label{eq7}
\end{equation}%
Since both the linearized optomechanical-coupling strength $G(t)$ and the
normalized driving detuning $\Delta(t)$ in Hamiltonian~(\ref{eq7}) depend on
$\alpha(t)$ and $\beta(t)$, the dynamic evolution of the system can be
controlled by adjusting the amplitude and phase of the pulsed driving, as
shown by Eqs.~(\ref{eq2a}) and (\ref{eq2b}).

The dynamic properties of the linearized optomechanical system are
completely described by both the first- and second-order moments of the
system operators. Using the relation $\partial _{t}\langle o_{i}o_{j}\rangle
=\text{Tr}(\dot{\rho}^{\prime }o_{i}o_{j})$ for $o_{i},o_{j}$ $\in $ \{$a$, $%
a^{\dagger }$, $b$, $b^{\dagger }$\} and Eq.~(\ref{Eq45}) under the
replacement of $H_{\mathrm{dis}}(t)\rightarrow H_{\mathrm{lin}}(t)$, we
obtain the equations of motion of all these second-order moments, which can be expressed as
\begin{equation}
\dot{\mathbf{X}}(t)=\mathbf{M}(t)\mathbf{X}(t)+\mathbf{N}(t),  \label{eq5.2}
\end{equation}%
where $\mathbf{X}(t)=(\langle a^{\dagger }a\rangle ,\langle b^{\dagger
}b\rangle ,\langle a^{\dagger }b\rangle ,\langle ab^{\dagger }\rangle
,\langle a^{\dagger }a^{\dagger }\rangle ,\langle a^{\dagger }b^{\dagger
}\rangle ,$ $\langle b^{\dagger }b^{\dagger }\rangle ,\langle aa\rangle
,\langle ab\rangle ,\langle bb\rangle )^{\text{T}}$ (\textquotedblleft $%
\mathrm{T}$" denotes the matrix transpose), $\mathbf{N}(t)=(0,\gamma \bar{n}_{\mathrm{m}%
},0,0,0,ig_{0}\alpha ^{\ast },0,0,-ig_{0}\alpha ,0)^{\mathrm{T}}$, and the coefficient matrix is
introduced as $\mathbf{M}(t)=\left(
\begin{array}{cc}
\mathbf{H }& \mathbf{I} \\
\mathbf{J} & \mathbf{K}%
\end{array}%
\right) $, with
\begin{subequations}\label{A9}
\begin{align}
\mathbf{H }&=\left(
\begin{array}{cccc}
-\kappa & 0 & -ig_{0}\alpha & ig_{0}\alpha ^{\ast }\\
0 & -\gamma &  ig_{0}\alpha & -ig_{0}\alpha ^{\ast }\\
-ig_{0}\alpha ^{\ast } & ig_{0}\alpha ^{\ast } & K_{1} & 0\\
ig_{0}\alpha & -ig_{0}\alpha & 0 & K_{1}^{\ast }
\end{array}%
\right), \\
\vspace{2em}
\mathbf{I }&=\left(
\begin{array}{cccccc}
 0 &  -ig_{0}\alpha &0 & 0 & ig_{0}\alpha ^{\ast } &0 \\
0 & -ig_{0}\alpha & 0 & 0 &ig_{0}\alpha ^{\ast } &0 \\
-ig_{0}\alpha & 0 & 0 & 0 & 0 & ig_{0}\alpha ^{\ast }\\
 0 &0 &  -ig_{0}\alpha  &  ig_{0}\alpha ^{\ast } & 0 &0
\end{array}
\right), \\
\vspace{2em}
\mathbf{J }&=\left(
\begin{array}{cccccc}
0 & 0 & 2ig_{0}\alpha ^{\ast } & 0 \\
ig_{0}\alpha ^{\ast } & ig_{0}\alpha ^{\ast } &0 & 0\\
0 & 0 & 0 & 2ig_{0}\alpha ^{\ast }\\
0  & 0 & 0 & -2ig_{0}\alpha \\
-ig_{0}\alpha & -ig_{0}\alpha & 0 & 0\\
0 & 0 & -2ig_{0}\alpha & 0
\end{array}
\right), \\
\vspace{2em}
\mathbf{K }&=\left(
\begin{array}{cccccc}
K_{2} & 2ig_{0}\alpha ^{\ast }
&  0 & 0 & 0 & 0 \\
ig_{0}\alpha & K_{3}  & ig_{0}\alpha ^{\ast } &  0 & 0 & 0 \\
0 & 2ig_{0}\alpha  & K_{4} & 0 &0 & 0 \\
0 &0 & 0 & K_{2} ^{\ast } & -2ig_{0}\alpha& 0\\
 0 & 0 & 0 &-ig_{0}\alpha  ^{\ast }& K_{3}  ^{\ast }& 0 \\
 0 & 0 & 0 & 0 & -2ig_{0}\alpha^{\ast } & K_{4}^{\ast }
\end{array}
\right).
\end{align}
\end{subequations}
 In Eq.~(\ref{A9}), we introduce
\begin{subequations}
\begin{align}
K_{1} &=i\Delta (t)-i\omega _{\mathrm{m}}-\frac{1}{2}(\kappa +\gamma ), \\
K_{2} &=2i\Delta (t)-\kappa , \\
K_{3} &=i\Delta (t)+i\omega _{\mathrm{m}}-\frac{1}{2}(\kappa +\gamma )\, \\
K_{4} &=2i\omega _{\mathrm{m}}-\gamma .
\end{align}%
\end{subequations}

We point out that both the coefficient matrix $\mathbf{M}(t)$ and the inhomogeneous term $\mathbf{N}%
(t)$ are functions of the displacement amplitudes $\alpha (t)$ and $\beta (t)
$, thus the dynamic evolution of the second-order moments can be controlled
via adjusting the pulsed driving.
\section{GENERATION OF MECHANICAL QUADRATURE SQUEEZING\label{13}}
\begin{table}[t]\label{algorithm}
\centering
\begin{tabular}{p{\dimexpr\linewidth-2\tabcolsep}}
\multicolumn{1}{c}{\textbf{Algorithm1.} Mechanical squeezing generation} \\
\hline
\hline
\textbf{Input:} A randomly smooth and continuous initial pulsed driving with
amplitude $\Omega$ and phase $\phi$. \\
\textbf{Define:} The loss function $\Delta X_{b}^{2}(\theta ,T)$. \\
\textbf{While} $\Delta X_{b}^{2}(\theta ,T)>\epsilon$ ($\epsilon$ is the
expected value) \textbf{do}: \\
\hspace{1em}1.~Perform the gradient-descent algorithm $\mathcal{Q}_{m+1}\!=\!%
\mathcal{Q}_{m}-\chi_{\mathcal{Q}}\{\delta [\Delta X_{b}^{2}(\theta ,T)]
/\delta \mathcal{Q}_{m}$\} ($\mathcal{Q}$ represents either $\Omega$ or $\phi$%
, $\chi_{\mathcal{Q}}$ is the learning rate, and $m$ indicates the iteration
number) to iteratively minimize $\Delta X_{b}^{2}(\theta ,T)$. \\
\hspace{1em}2. Optimize and update $\Omega$ and $\phi$. \\
\textbf{End While} \\
\textbf{Return:} Variance $\Delta X_{b}^{2}(\theta ,t)$, driving
amplitude $\Omega$, and phase $\phi$. \\ \hline
\hline
\end{tabular}
\end{table}
To quantify the created mechanical squeezing, we introduce
the quadrature operators $%
X_{o=a,b}=(o^{\dagger }+o)/\sqrt{2}$ and $Y_{o=a,b}=i(o^{\dagger }-o)/\sqrt{2%
}$ for the optical and mechanical modes. Then the transient correlation matrix can be introduced as
\begin{equation}
\mathbf{V}_{ij}(t)=\frac{1}{2}[\langle \mathbf{u}_{i}(t)\mathbf{u}%
_{j}(t)\rangle +\langle \mathbf{u}_{j}(t)\mathbf{u}_{i}(t)\rangle ]+\langle
\mathbf{u}_{i}(t)\rangle \langle \mathbf{u}_{j}(t)\rangle,
\end{equation}
where $\mathbf{u}(t)=(X_{a}(t),Y_{a}(t),X_{b}(t),Y_{b}(t))^{\mathrm{T}}$.~In
the present case, the expectation values $\langle \mathbf{u}_{i}(t)\rangle $
and $\langle \mathbf{u}_{j}(t)\rangle $ are zero and $\langle \mathbf{u}%
_{i}(t)\mathbf{u}_{j}(t)\rangle$ is a linear function of these second-order
moments. To describe the quadrature squeezing, we further introduce the
rotating-quadrature operator $X_{b}(\theta ,t)\equiv X_{b}(t)\cos \theta
+Y_{b}(t)\sin \theta $ ($\theta$ is the squeezing angle). The variance of
the rotating-quadrature operator is given by
\begin{eqnarray}  \label{eq12}
\Delta X_{b}^{2}(\theta ,t) &=\mathbf{V}_{33}(t)\cos ^{2}\theta +\mathbf{V}%
_{44}(t)\sin ^{2}\theta \notag\\ & +\frac{1}{2}[\mathbf{V}_{34}(t)+\mathbf{V}_{43}(t)]\sin (2\theta).
\end{eqnarray}
Based on $[X_{b}(\theta ,t),X_{b}(\theta +\pi ,t)]\!=\!i$, we have $\Delta
X_{b}^{2}(\theta ,t) \Delta X_{b}^{2}(\theta+\pi ,t) \geq 1/4$, then the
quadrature squeezing of the mechanical mode occurs when $\Delta
X_{b}^{2}(\theta ,t) <1/2$.
\begin{figure}[bp]
\centering \includegraphics[width=0.48\textwidth]{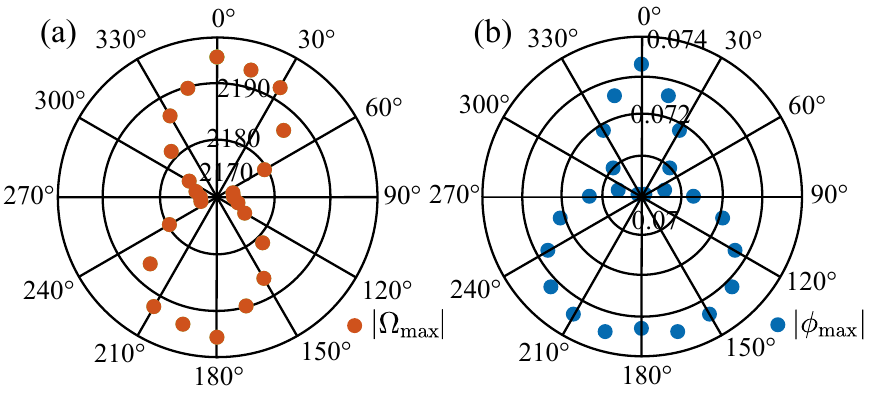}
\caption{Maximal absolute values of (a) the driving amplitude and (b) phase as functions of squeezing angle $\protect\theta$ when the squeezing degree $%
\mathcal{S}_{b}=1$. Here, the parameters used are $g_{0}/\protect\omega_{\mathrm{m%
}}=4\times10^{-5}$, $\protect\kappa/\protect\omega_{\mathrm{m}}=0.2$, $%
\protect\gamma/\protect\omega_{\mathrm{m}}=2\times10^{-6}$, $T=120\protect%
\omega_{\mathrm{m}}^{-1}$, $\bar{n}_{\mathrm{m}}$=100, and $\Delta_{\mathrm{c%
}}/\protect\omega_{\mathrm{m}}=1 $.}
\label{Fig1}
\end{figure}
For the squeezing-generation task, our goal is to reduce the variance $\Delta
X_{b}^{2}(\theta ,t)$ such that it is smaller than 1/2. The control strategy
of squeezing generation is summarized in $\mathbf{Algorithm1}$. Note that a variable learning rate method is adopted here to reduce the
iteration number and to improve the efficiency. The detailed calculation of $%
\delta [\Delta X_{b}^{2}(\theta ,T)]/\delta \mathcal{Q}_{m}$ is shown in
Appendix.

\begin{figure*}[tbp]
\centering \includegraphics[width=0.98\textwidth]{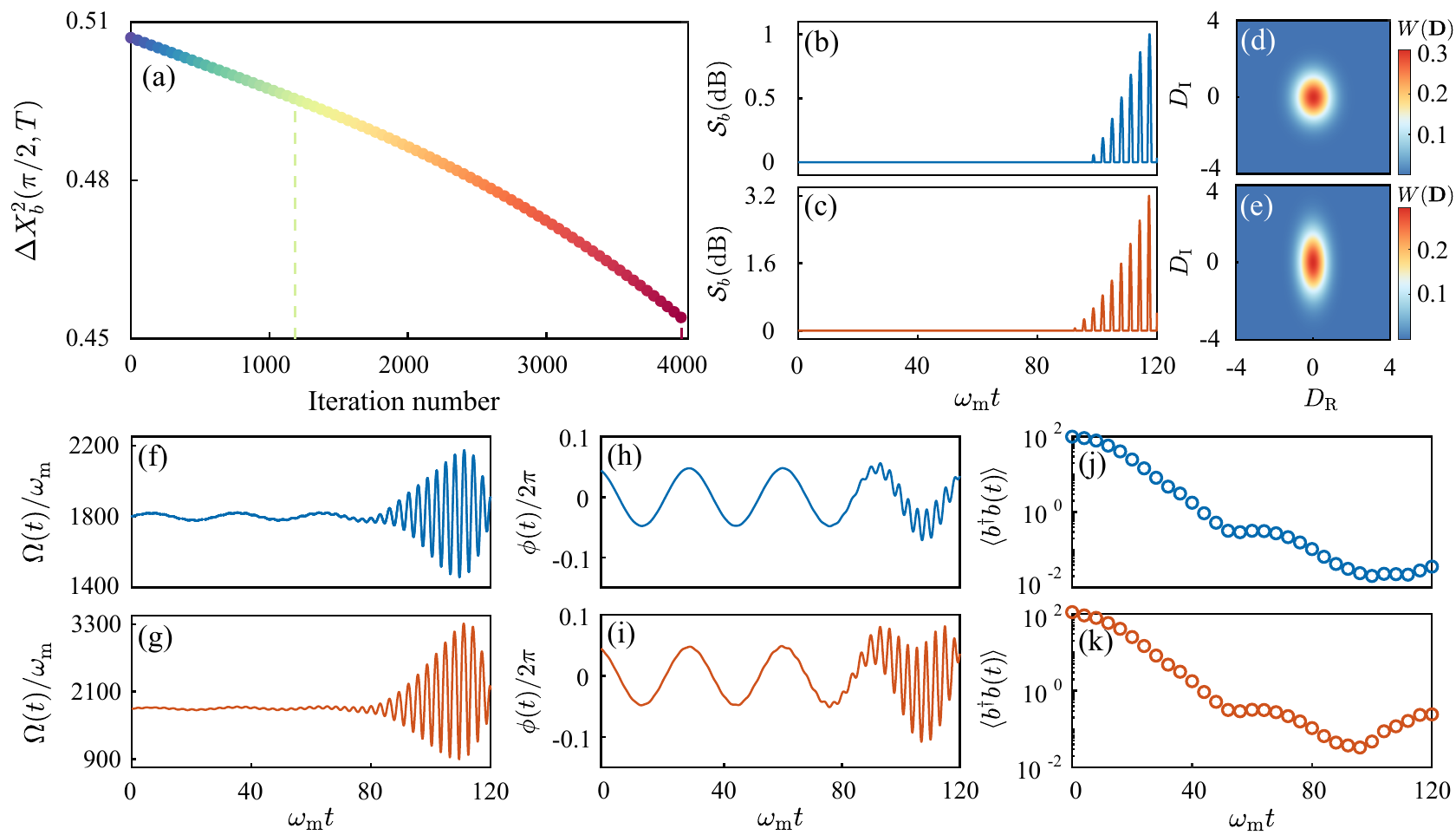}
\caption{(a) The loss function $\Delta X_{b}^{2}(\protect\theta=\protect\pi%
/2 ,T)$ as a function of the iteration number. The squeezing degree $\mathcal{S}_{b}$
vs the scaled evolution time $\protect\omega_{\mathrm{m}}t$ under
different target squeezing degrees: (b) $\mathcal{S}_{b}=1$ and (c) $%
\mathcal{S}_{b}=3.2$. (d),(e) The Wigner functions of the generated states for the
mechanical mode corresponding to the maximal squeezing in panels (b) and (c).
(f),(g) The scaled driving amplitude $\Omega(t)/\protect\omega_{\mathrm{m}}$, (h),(i)
phase $\protect\phi(t)/2\protect\pi$, and (j),(k) mean phonon number $\langle
b^{\dagger}b\rangle$ as functions of the scaled evolution time $\protect%
\omega_{\mathrm{m}}t$ corresponding to panels (b) and (c). Other parameters used are
the same as those given in Fig.~\protect\ref{Fig1}.}
\label{Fig2}
\end{figure*}
The mechanical quadrature squeezing can be well quantized by the squeezing
degree~\cite{AgarwalPRA2016}
\begin{equation}
\mathcal{S}_{b}=-10\log_{10}\frac{\Delta X_{b}^{2}( \theta,t) }{\Delta
X_{b}^{2}( \theta)|_{\mathrm{zpf}}} .
\end{equation}
A positive squeezing degree, $\mathcal{S}_{b}>0$, implies that the
quadrature operator of the mechanical mode is squeezed along the angle $%
\theta$. For studying the rotating-quadrature squeezing, we need to investigate the dependence of
the squeezing on the angle $\theta$. Concretely, we use the
gradient-descent algorithm to obtain the waveforms of the driving
amplitude and phase corresponding to $\mathcal{S}_{b}=1$ at different values
of $\theta$. The results shown in Fig.~\ref{Fig1}(a) indicate that the
smallest (largest) driving amplitude appears at the squeezing angle $%
90^{\circ}$ ($0^{\circ}$) related to $\mathcal{S}_{b}=1$. Similarly, the
minimal and maximal driving phases appear at $65^{\circ}$ and $165^{\circ}$,
respectively, and the angle has little effect on the phase [as shown in Fig.~%
\ref{Fig1}(b)]. Therefore, we take the squeezing angle $\theta=90^{\circ}$
(corresponding to minimal driving amplitude) in our following discussions.

We display in Fig.~\ref{Fig2}(a) the loss function $\Delta X_{b}^{2}(\pi/2
,T)$ as a function of the iteration number to verify the efficiency of the
gradient-descent algorithm. Here, we can see that the loss function $%
\Delta X_{b}^{2}(\pi/2 ,T)$ decreases gradually as the iteration number
increases. The transient mechanical squeezing degrees $1$ and $3.2$ dB are
obtained at the iteration numbers $1185$ and $4000$, respectively. The corresponding
dynamic evolutions of the squeezing degree $\mathcal{S}_{b}$ are shown in
Figs.~\ref{Fig2}(b) and~\ref{Fig2}(c). It shows that there is no mechanical
squeezing for a long duration of time, first, and then oscillation
increases to the target value in a short period of time. In particular, the
squeezing degree $\mathcal{S}_{b}=3.2$ breaks the $3$-dB steady-state limit
[Fig.~\ref{Fig2}(c)], and stronger squeezings can be realized as the iteration
number increases. Note that we consider the transient squeezing here rather
than steady-state squeezing. Therefore, our results are not conflicting with
the $3$-dB steady-state squeezing limit, which is determined by the dynamic stability.

To confirm the mechanical squeezing in phase space, we introduce the Wigner
function of the mechanical mode, defined as~\cite{Vogel2006}
\begin{equation}
W(\mathbf{D})=\frac{1}{2\pi \sqrt{\mathrm{Det}[\mathbf{V}_{b}]}}\exp \left\{-%
\frac{1}{2}\mathbf{D}^{\mathrm{T}}\mathbf{V}_{b}\mathbf{D}\right\},
\end{equation}
where $\mathbf{D}=(D_{\mathrm{R}},D_{\mathrm{I}})^{\mathrm{T}}$ represents
the two-dimensional vector, and $\mathbf{V}_{b}$ is the covariance matrix
for the mechanical mode. The Wigner functions corresponding to 1- and 3.2-dB squeezing are shown in Figs.~\ref{Fig2}(d) and~\ref{Fig2}(e). We can see
that the squeezing appears along the $\pi/2$ axis (corresponding to the
squeezing angle $\theta=\pi/2$), and the larger the squeezing degree the
stronger the quadrature squeezing.
\begin{figure}[tbp]
\centering \includegraphics[width=0.46\textwidth]{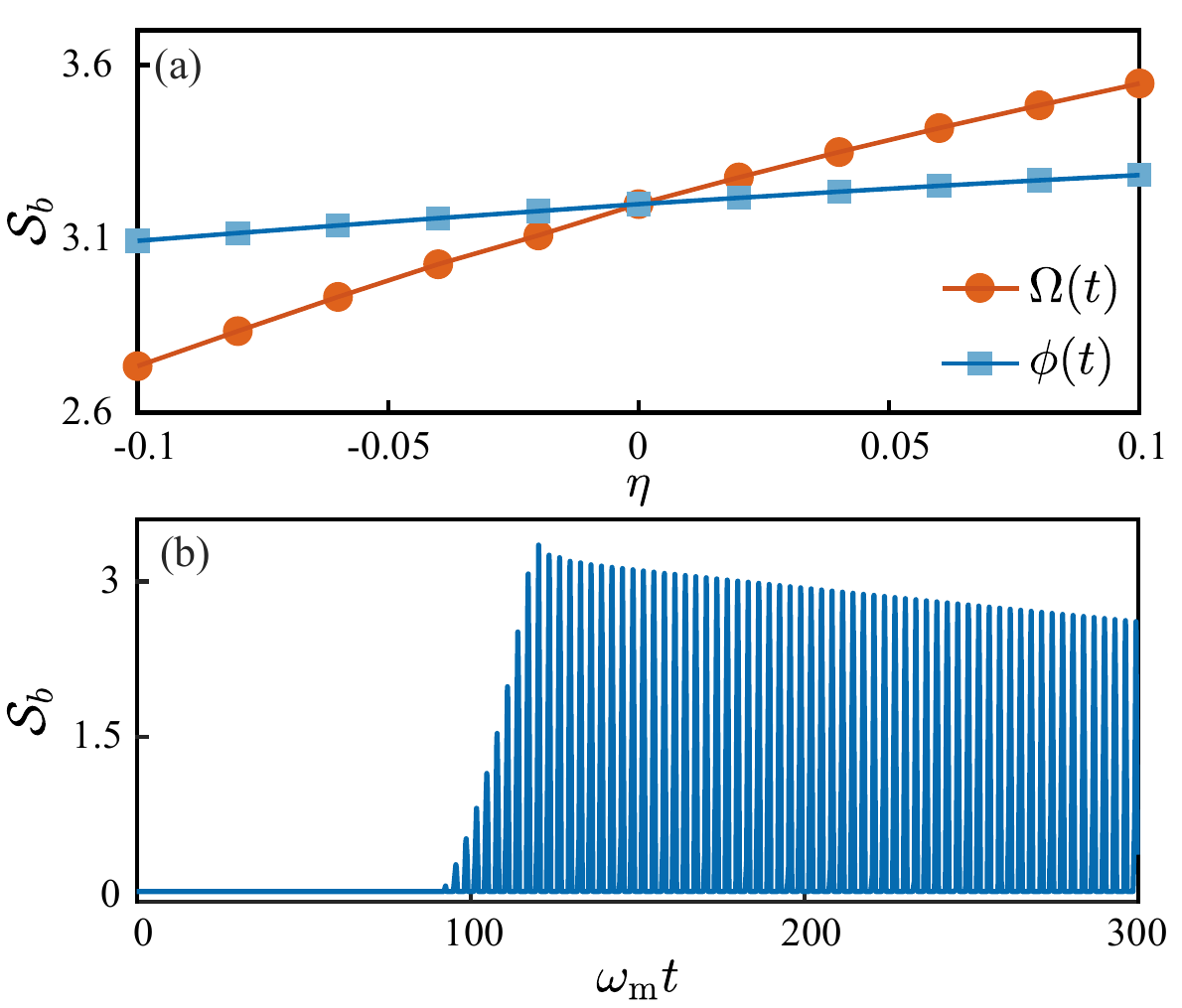}
\caption{(a) The influence of the driving amplitude and phase deviations on
the squeezing generation at $T=120\protect\omega_{\mathrm{m}}^{-1}$ related
to $\mathcal{S}_{b}=3.2$. (b) The dynamic evolution of the squeezing degree $%
\mathcal{S}_{b}$ after removing the pulsed driving in Fig.~\protect\ref{Fig2}%
(c). Other parameters used are consistent with those given in Fig.~\protect\ref{Fig1}.}
\label{Fig3}
\end{figure}

In Figs.~\ref{Fig2}(f) and~\ref{Fig2}(g), we show the time-dependent driving
amplitudes, which are required to achieve the squeezing degrees $\mathcal{S}%
_{b}=1$ and 3.2 dB, respectively~\cite{check}. We can see
that a larger driving amplitude is required to realize a stronger squeezing.
In addition, for a larger mechanical squeezing degree, the driving phase
oscillation becomes more intense, as shown in Figs.~\ref{Fig2}(h)
and~\ref{Fig2}(i). We also investigate the
dependence of the mean phonon number $\langle b^{\dagger}b\rangle$ (in the displaced representation, associated with the quantum fluctuation) as a function of the scaled time $\omega_{\mathrm{m}}t$ in Figs.~\ref{Fig2}(j) and~\ref{Fig2}%
(k). We observe that $\langle b^{\dagger}b\rangle$ decreases from 100 to less than 1 for both $\mathcal{S}_{b}=1$ and 3.2 dB. Note that in our simulations, the mechanical resonator is assumed initially in a thermal state at the same temperature with the heat bath. The density matrix for the thermal state is given by $\rho_{\mathrm{th}}=\sum_{n=0}^{\infty}P_{n}|n\rangle\langle n|$, where $P_{n}=\bar{n}_{\mathrm{th}}^{n}/(\bar{n}_{\mathrm{th}}+1)^{n+1}$ represents the probability distribution, with the mean thermal phonon number $\bar{n}_{\mathrm{th}}=\mathrm{Tr}(\rho_{\mathrm{th}}b^{\dagger}b)=1/(e^{\hbar\omega_{\mathrm{m}}/k_{B}T}-1)$.

\begin{figure}[tbp]
\centering \includegraphics[width=0.46\textwidth]{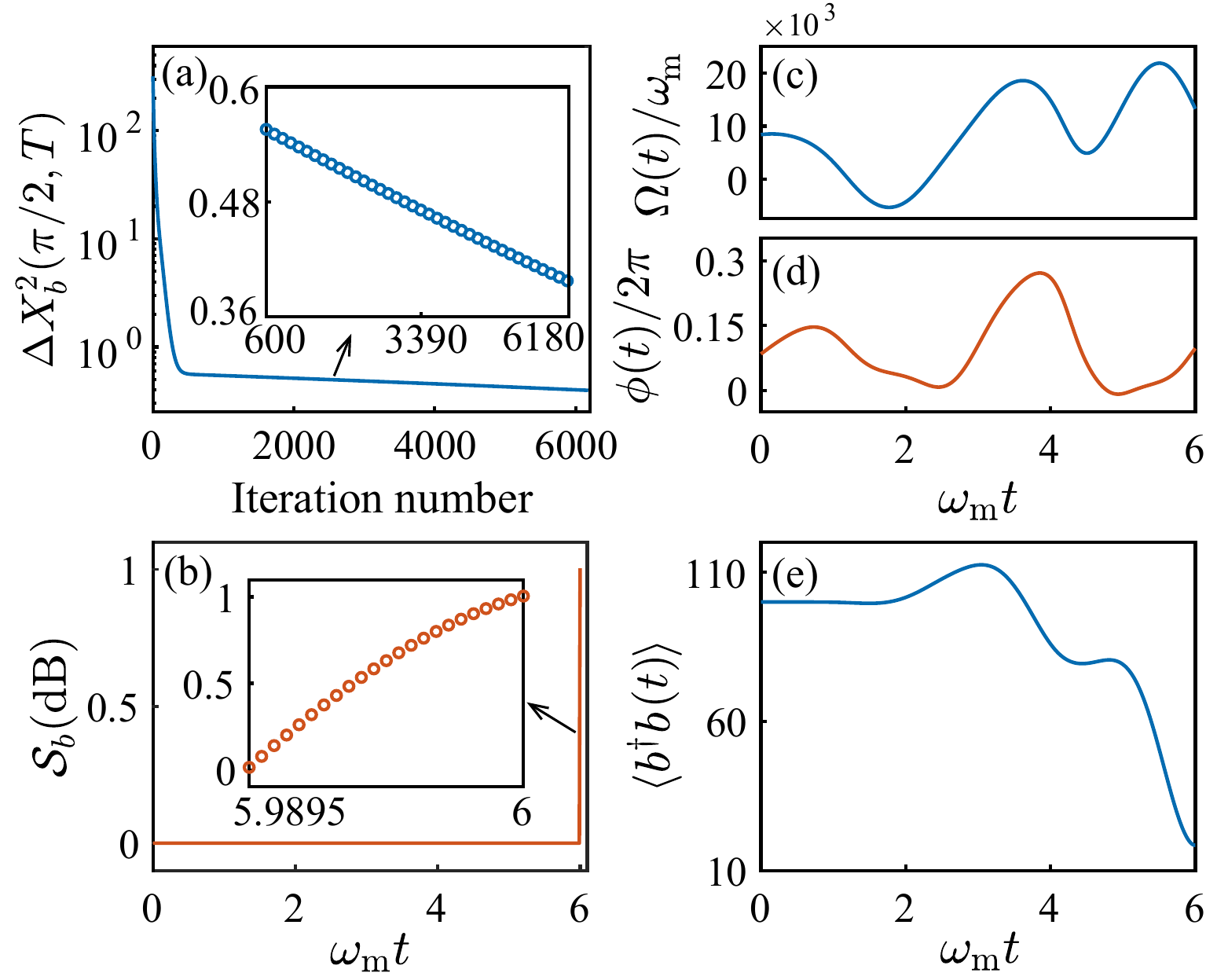}
\caption{(a) The loss function $\Delta X_{b}^{2}( \protect\pi/2,T)$ as a
function of the iteration number under $T=6\protect\omega_{\mathrm{m}}^{-1}$. The
inset shows the local magnification of the loss function over the iteration
interval [600,6180]. (b) Evolution of the squeezing degree $\mathcal{S}_{b}$
in one mechanical oscillation period. The inset shows the evolution of $%
\mathcal{S}_{b}$ over time duration [5.9895,6]. (c)~The driving amplitude $%
\Omega(t)/\protect\omega_{\mathrm{m}}$ and (d) phase $\protect\phi(t)/2%
\protect\pi$ vs the scaled evolution time $\protect\omega_{\mathrm{m}}t$
after the last iteration ($m=6180$) under $T=6\protect\omega_{\mathrm{m}%
}^{-1}$. (e) The mean phonon number $\langle b^{\dagger}b\rangle$ as a
function of the scaled evolution time $\protect\omega_{\mathrm{m}}t$
corresponding to panel (b). Other parameters used are the same as those given in Fig.~\protect\ref%
{Fig1}.}
\label{Fig4}
\end{figure}

To explore the influence of the deviations in $\Omega(t)$ and $\phi(t)$ on
mechanical quadrature squeezing, we introduce the relative deviation $\eta$,
defined as $\eta=(Q_{r}-Q_{t})/Q_{t}$. Here $Q$ represents either $\Omega$
or $\phi$, and $Q_{r}$ and $Q_{t}$ stand for the realistically used parameters and
the learned theoretical parameters, respectively. In Fig.~\ref{Fig3}(a), we
plot the squeezing degree $\mathcal{S}_{b}$ as a function of $\eta$. Here,
we see that for the relative deviation $\eta\in[-0.1,0.1]$, the $\mathcal{S}%
_{b}$ is an increasing function of $\eta$ for both the driving amplitude and phase. In
addition, the squeezing is more sensitive to amplitude deviation than phase
deviation. In Fig.~\ref{Fig3}(b), we further investigate the influence of
the environment on the squeezing generation once the pulsed driving is
removed at $T=120\omega_{\mathrm{m}}^{-1}$. The $\mathcal{S}_{b}$ will
experience a slight decrease from 3.2 to 2.51 during the period from $%
\omega_{\mathrm{m}}t=120$ to 300, indicating that the squeezing has a good
robustness. Notably, after the external drive is removed, the photons in the cavity will dissipate completely within a certain period, and the optomechanical interaction will no
longer work. Consequently, the mechanical resonator will be thermalized by  the heat bath, leading to a reduction of the mechanical quadrature squeezing.

\begin{figure}[tbp]
\centering \includegraphics[width=0.48\textwidth]{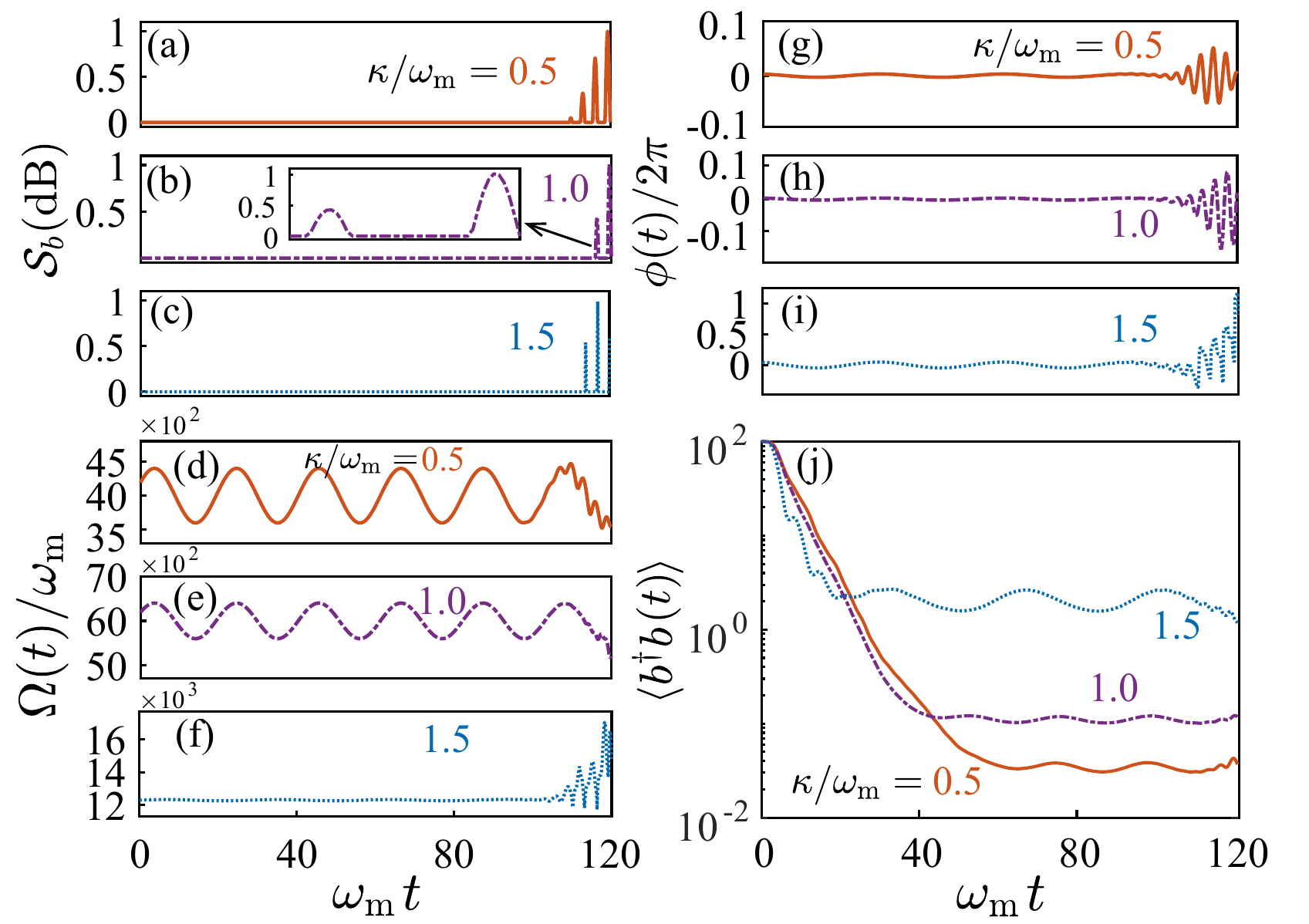}
\caption{(a)-(c) The squeezing degree $\mathcal{S}_{b}$ vs the scaled
evolution time $\protect\omega_{\mathrm{m}}t$ under different sideband
parameters $\protect\kappa/\protect\omega_{\mathrm{m}}=0.5$, 1.0, and 1.5.
The inset in panel (b) exhibits the evolution of $\mathcal{S}_{b}$ for the
time durations from $115.6$ to $120$. (d)-(f) Evolution of
driving amplitude and (g)-(i) phase corresponding to panels (a)-(c). (j) The dynamic evolution of the mean
phonon number under different sideband-resolution parameters in panels (a)-(c). The two insets show the enlarged view at the peak of 3.2 dB. Other parameters are consistent with those given in Fig.~\protect\ref{Fig1}.}
\label{Fig5}
\end{figure}
The ultrafast generation of mechanical quadrature squeezing within one mechanical oscillation period is a desired task from the
viewpoint of transient evolution~\cite{TrianaPRL2016,LiuPRA2024}. Below, we investigate the generation of
mechanical squeezing within this short timescale. Figure~\ref%
{Fig4}(a) shows the loss function $\Delta X_{b}^{2}( \pi/2,T=6\omega_{%
\mathrm{m}}^{-1})$ versus the iteration number related to $\mathcal{S}_{b}=1
$. Here we can see that the $\Delta X_{b}^{2}( \pi/2,T=6\omega_{\mathrm{m}%
}^{-1})$ gradually decreases from the initial value 319 to 0.397 as the
iteration number increases, verifying the validity of the gradient-descent algorithm. The dynamic evolution of the squeezing degree $\mathcal{S}%
_{b}$ after the last iteration ($m=6180$) is shown in Fig.~\ref{Fig4}(b). We
see that the squeezing occurs in the last extremely short duration due to
the harsh time condition, but it is still continuous, as shown in the inset
of Fig.~\ref{Fig4}(b). The results in Figs.~\ref{Fig4}(c) and~\ref{Fig4}(d)
indicate that a large driving amplitude is required to achieve $\mathcal{S}%
_{b}=1$ within one mechanical oscillation period and the phase has no
distinct signature. We also exhibit in Fig.~\ref{Fig4}(e) the mean phonon
number $\langle b^{\dagger}b\rangle$ versus the scaled evolution time at
this time. Here we see that $\langle b^{\dagger}b\rangle$ will be larger
than the initial value due to the larger driving amplitude, and then goes
to a cooled state with dozens of mean phonons. This indicates that the
creation of mechanical squeezing does not necessarily require the
ground-state cooling of the mechanical resonator.

We also investigate the influence of the sideband-resolution parameter on
the squeezing generation. In Figs.~\ref{Fig5}(a) to~\ref{Fig5}(c), we plot the squeezing
degree $\mathcal{S}_{b}$ versus the scaled evolution time $\omega_{\mathrm{m}%
}t$ corresponding to different sideband-resolution parameters $\kappa/\omega_{\mathrm{m}%
}=0.5$, 1.0, and 1.5. We show the driving amplitude and phase as functions of $%
\omega_{\mathrm{m}}t$ for different sideband-resolution parameters in Figs.~\ref{Fig5}%
(d)-\ref{Fig5}(f) and Figs.~\ref{Fig5}(g)-\ref{Fig5}(i). To maintain $\mathcal{S}_{b}=1$, a larger
driving amplitude is needed for a larger value of $\kappa/\omega_{%
\mathrm{m}}$, and the phase oscillation becomes more severe for a larger $%
\kappa/\omega_{\mathrm{m}}$. The corresponding dynamic evolution of the mean
phonon number $\langle b^{\dagger}b\rangle$ is shown in Fig.~\ref{Fig5}(j).
Here we can see that the mechanical resonator is cooled from the initial 100 phonons to a few phonons. The better the sideband-resolution parameter that is
selected, the deeper the cooling of the mechanical resonator. Note that a large $\kappa/\omega_{%
\mathrm{m}}$ will result in a fast dissipation of the cavity photons, thereby affecting the
optomechanical interaction and  leading to the failure of squeezing generation. We mention that the thermal excitations are usually harmful to the generation and maintenance of quantum signatures. For the generation of strong quadrature squeezing, optimal parameters should be chosen such that the thermal excitations can be decreased to a near-ground-state cooling level.

\begin{figure}[tbp]
\centering \includegraphics[width=0.48\textwidth]{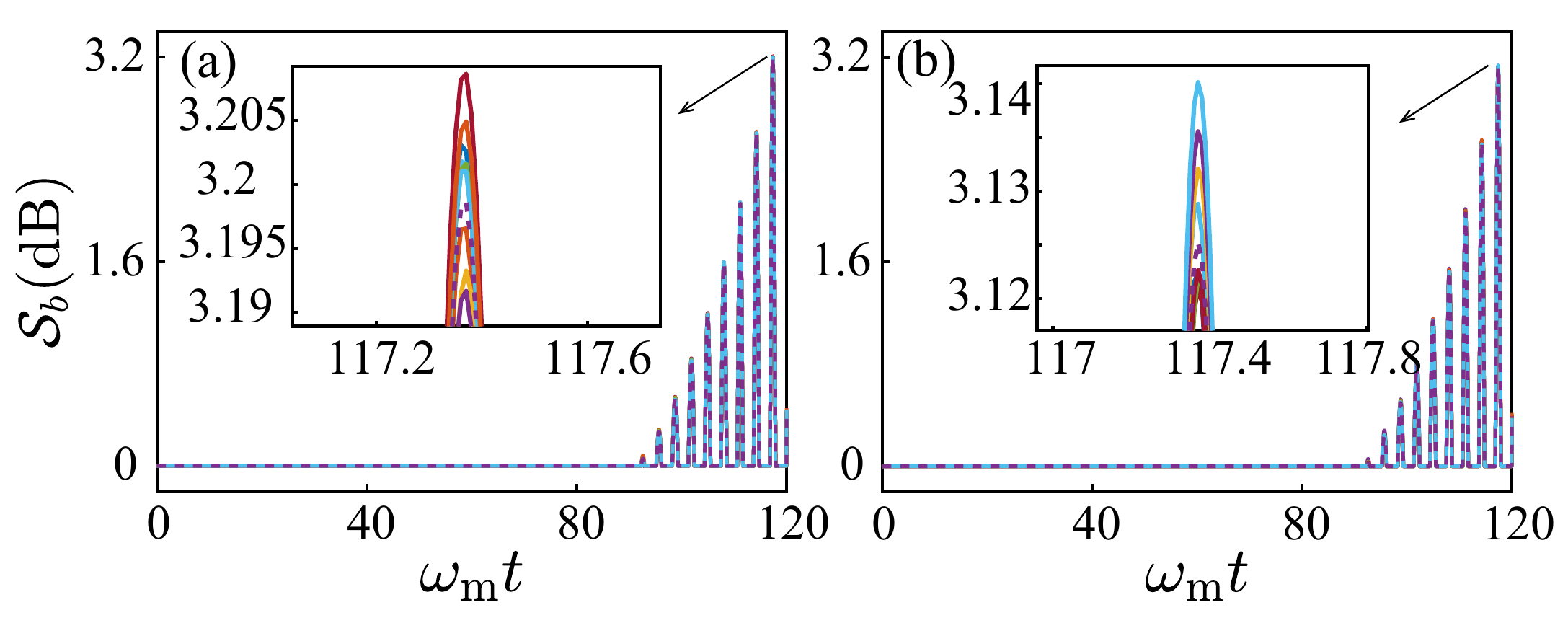}
\caption{The squeezing degree $\mathcal{S}_{b}$ versus the scaled evolution time $\omega_{\mathrm{m}}t$ in the presence of  (a) driving amplitude and (b) phase noises. The driving amplitude is modified by adding a scaled random amplitude from a normal distribution with a mean of 0 and a standard deviation of 200. Similarly, the driving phase is modified by adding a scaled random phase from a normal distribution with a mean of 0 and a standard deviation of 0.2. Other parameters are the same as those in Fig.~\ref{Fig2}(c).}
\label{Fig6}
\end{figure}
\section{DISCUSSIONS\label{14}}
In this section, we present some discussions concerning the influences of the driving amplitude and phase noises on the squeezing generation, the physical mechanism of the present squeezing-generation method, the experimental implementation of this scheme, and the applications of the optimization methods in cavity optomechanics.

\subsection{Influences of the driving amplitude and phase noises on squeezing generation}
In realistic situations, the driving laser inevitably possesses amplitude and phase noises, which can be understood as random small deviations on the amplitude and phase.  To evaluate the influences of laser amplitude and phase noises on the squeezing-generation performance, we add Gaussian random noises to the optimized pulse driving amplitude and phase at each step. Figures~\ref{Fig6}(a) and~\ref{Fig6}(b) display the squeezing degree $\mathcal{S}_{b}$  as a function of the scaled evolution time $\omega_{\mathrm{m}}t$ after introducing the Gaussian random noises to the driving amplitude and phase, respectively. The ten solid curves represent the evolution of the squeezing degree $\mathcal{S}_{b}$ under ten distinct random conditions, while the purple dotted curve represents the average of these ten evolutions. We can see that the 11 curves almost overlap, which indicates that the optimized driving field is stable against both the amplitude and phase noises.

\subsection{Physical mechanism of the present squeezing-generation method}
In our scheme, the generation of mechanical squeezing can be understood by analyzing the variance of the mechanical rotating-quadrature operator $\Delta
X_{b}^{2}(\theta ,t) =\mathbf{V}_{33}(t)\cos ^{2}\theta +\mathbf{V}%
_{44}(t)\sin ^{2}\theta +\frac{1}{2}[\mathbf{V}_{34}(t)+\mathbf{V}%
_{43}(t)]\sin (2\theta)$. To clarify this point, we re-express the rotating-quadrature operator as
\begin{align}
\Delta X_{b}^{2}(\theta ,t) &=\frac{1}{2}+\langle b^{\dagger}b\rangle+%
\mathrm{Re}[\langle b^{\dagger}b^{\dagger}\rangle]\cos(2\theta)
\notag \\  &\hspace{0.3cm}-2\mathrm{Im}[\langle b^{\dagger}b^{\dagger}\rangle]\sin(2\theta).
\end{align}
Mechanical squeezing occurs when $\Delta X_{b}^{2}(\theta ,t) <1/2$, which leads to the condition
\begin{equation}
  \langle b^{\dagger}b\rangle+\mathrm{Re}%
[\langle b^{\dagger}b^{\dagger}\rangle]\cos(2\theta)-2\mathrm{Im}[\langle
b^{\dagger}b^{\dagger}\rangle]\sin(2\theta)<0.
\end{equation}
This inequality can be satisfied because the second-order moment $\langle b^{\dagger}b^{\dagger}\rangle$ could be a
complex number, while $\langle b^{\dagger}b\rangle$ must
be positive. Our numerical simulations indicate the mechanical cooling is accompanied by the physics process, and the mechanical resonator is cooled to a relatively small mean phonon number [Figs.~\ref{Fig2}(j),~\ref{Fig2}(k),~\ref{Fig4}(e), and~%
\ref{Fig5}(j)]. Therefore, the generation of the mechanical squeezing is primarily determined by the values of $\mathrm{Re}[\langle b^{\dagger}b^{\dagger}\rangle]$ and $\mathrm{Im}[\langle b^{\dagger}b^{\dagger}\rangle]$.

Under the rotating-wave approximation, the second-order moments are governed by two separate sets of equations. The first set includes $\langle a^{\dagger}a\rangle$, $\langle
b^{\dagger}b\rangle$, $\langle a^{\dagger}b\rangle$, and $\langle
ab^{\dagger}\rangle$, while the second set includes $\langle
a^{\dagger}a^{\dagger}\rangle$, $\langle aa\rangle$, $\langle
b^{\dagger}b^{\dagger}\rangle$, $\langle bb\rangle$, $\langle
a^{\dagger}b^{\dagger}\rangle$, and $\langle ab\rangle$. With the initial
condition $\mathbf{X}(0)=[0,\bar{n}_{\mathrm{m}},0,0,0,0,0,0,0,0]^{\mathrm{T}}$, the value
of $\langle b^{\dagger}b^{\dagger}\rangle$ remains constantly zero,
leading to a negligible mechanical squeezing regardless of how we optimize the
driving fields. The counter-rotating terms in the coupling will mix the equations of motion for these ten second-order moments. In
this case, we can obtain $\langle b^{\dagger}b\rangle+\mathrm{Re}[\langle
b^{\dagger}b^{\dagger}\rangle]\cos(2\theta)-2\mathrm{Im}[\langle
b^{\dagger}b^{\dagger}\rangle]\sin(2\theta)<0$ via optimizing the pulsed
driving, indicating the generation of squeezing. These analyses indicate that the critical role of the counter-rotating terms is the generation of mechanical quadrature squeezing.

Meanwhile, we examined the single-mode squeezing properties of the two-mode squeezed vacuum state and the state obtained by applying the two-mode squeezing operator to the direct product of vacuum and thermal states. We find that there is no single-mode quadrature squeezing in these two cases, indicating that the pure counter-rotating-wave terms are insufficient for generating single-mode squeezing. Based on these analyses, we deduce that the generation of single-mode quadrature squeezing arises from the combined effects of both rotating-wave and counter-rotating-wave terms. Since both the linearized optomechanical-coupling strength $G(t)$ and the normalized driving detuning $\Delta(t)$ depend on the evolution time, these provide the possibility to induce the physical process for effective two-phonon interactions, which is the physical origin for generating single-mode squeezing.

\subsection{Experimental implementation of the scheme}
We now present some discussions on the experimental feasibility of this scheme. The system under considerations is a general optomechanical
system, which has been implemented in various optomechanical platforms, such
as optical microresonators~\cite%
{ParkNP2009,SchliesserNP2009,VerhagenNature2012}, electromechanical systems~%
\cite{HertzbergNP2010,MasselN2011,MasselNC2012}, photonic crystal nanobeams~\cite%
{TeufelNature2011,Eichenfield2009,ChanNT2009}, and Fabry-P\'{e}rot cavities~\cite%
{HammererNature2009}, in which the cavities can be driven
by tunable pulsed fields. We considered the linearized optomechanical
interaction, which has been widely demonstrated in cavity optomechanical systems. Moreover, we use the experimentally accessible parameters in our numerical simulations. Concretely,
the used parameters are $g_{0}/\omega_{\mathrm{m}}=4\times10^{-5}$, $%
\kappa/\omega_{\mathrm{m}}=0.2$, and $\gamma/\omega_{\mathrm{m}%
}=2\times10^{-6}$, which have been reported in experiments~\cite%
{VerhagenNature2012}. In addition, the pulsed driving amplitude and phase
used to generate squeezing are moderate in size and
continuous and smooth in shape, which confirm the experimental realization
of the pulsed driving.

\subsection{Applications of the optimization methods in \\ cavity optomechanics}
To broaden the use of our optimization method in the field of cavity optomechanics, we investigate its applicability across two primary research regimes: the strong driving linearization regime and the single-photon strong-coupling regime.

In the strong driving linearization regime, the physical properties of the optomechanical system are governed by the covariance matrix. For this linearized optomechanical system, we can control the dynamics of the system by optimizing the driving field, which is an adjustable control parameter. For example, we can optimize the physical topics such as optomechanical cooling, entanglement, and squeezing. Moreover, this optimization problem can be extended and applied to multi-mode optomechanical systems.

In the single-photon strong-coupling regime, the optomechanical system can be described as a multilevel system in the eigen representation of the photon-number-dependent displaced oscillator system. The driving field will induce the transitions between the states associated with neighboring photon numbers. Here, the transition magnitudes depend on the external driving field, which provide the means for controlling the systems. Typically, the optimization allows for several potential applications, including the enhancement of the photon-blockade effect and the preparation of non-classical states such as mechanical number states. By implementing these optimizations, we can deepen both the understanding and the practical applications of the cavity optomechanical systems.

\section{CONCLUSION\label{15}}
In conclusion, we have presented a scheme for generating mechanical
quadrature squeezing in a typical optomechanical system via gradient-descent algorithm. The generated mechanical squeezing can
exceed the 3-dB steady-state limit and ultrafast squeezing preparation
within one mechanical oscillation period can be realized. The optimal
driving amplitude and phase corresponding to these generated squeezings have
been presented. Our scheme will pave the way for exploiting optimal quantum control in quantum optics and quantum information science.

\begin{acknowledgments}
The authors thank Dr. Ye-Xiong Zeng and Dr. Yue-Hui Zhou for helpful discussions. J.-Q.L. was supported in part by National Natural Science Foundation of China (Grants No.~12175061, No.~12247105, and No.~11935006), National Key Research and Development Program of China (Grant No.~2024YFE0102400), and Hunan Provincial Major Sci-Tech Program (Grant No.~2023ZJ1010). Y.-H.L. was supported in part by Hunan Provincial Postgraduate Research and Innovation project (Grant No.~CX20240530).
\end{acknowledgments}

\appendix*
\section{Derivation of the variation $\delta [\Delta X_{b}^{2}(\theta ,T)]/\delta \mathcal{Q}_{m}$\label{6678}}
\setcounter{section}{1}
\addcontentsline{toc}{section}{Appendix}
For generation of mechanical quadrature squeezing, we need to minimize the loss function $\Delta X_{b}^{2}(\theta ,T)$. In this Appendix, we will clarify in detail how to achieve this goal using the gradient-descent algorithm. The mathematical expression of the gradient-descent algorithm for mechanical squeezing generation is
\begin{equation}\label{eqb1}
\mathcal{Q}_{m+1}=\mathcal{Q}_{m}-\chi_{\mathcal{Q}}\frac{\delta [\Delta X_{b}^{2}(\theta ,T)]}{\delta\mathcal{Q}_{m}},
\end{equation}
where $\mathcal{Q}$ could be either $\Omega$ or $\phi$, $m$ is the iteration number, and $\chi_{\mathcal{Q}}$ is the learning rate. To calculate Eq.~(\ref{eqb1}), we need to calculate the variation of $\Delta X_{b}^{2}(\theta ,T) $ with respect to the pulse amplitude $\Omega(s)$ and phase $\phi(s)$. Based on Eq.~(\ref{eq12}), we can obtain the result
\begin{eqnarray}\small \label{subeq2}
\!\frac{\delta \Delta X_{b}^{2}(\theta ,T) }{\delta
\mathcal{Q}(s)}\! \!&=&\!\!\frac{\delta \mathrm{V}_{33}(T)}{\delta \mathcal{Q}(s)}%
\cos ^{2}\theta+\frac{\delta \mathrm{V}_{44}(T)}{\delta \mathcal{Q}(s)}\sin ^{2}\theta \notag\\ && +
\left[\frac{\delta \mathrm{V}_{34}(T)}{\delta \mathcal{Q}(s) }+\frac{\delta \mathrm{V}_{43}(T)}{\delta \mathcal{Q}(s) }\right]\sin (2\theta). 
\end{eqnarray}%
According to Eq.~(\ref{subeq2}), we can further calculate the results of $\delta
\mathrm{V}_{33}( T) /\delta\mathcal{Q} ( s) $, $\delta \mathrm{V}_{44}( T)
/\delta\mathcal{Q} ( s) $, $\delta \mathrm{V}_{34}( T) /\delta\mathcal{Q}
( s)$, and $\delta \mathrm{V}_{43}( T) /\delta\mathcal{Q}
( s)$. The variation of the covariance-matrix elements with respect
to the pulse amplitude $\Omega(s)$ and phase $\phi(s)$ can be expressed as a linear combination of the variation of these second-order moments with respect to the driving amplitude $\Omega(s)$ and phase $\phi(s)$:
\begin{widetext}
\begin{subequations}\label{A3}
\begin{align}
&\frac{\delta \mathbf{V}_{11}( T) }{\delta \mathcal{Q}( s)} =\frac{1}{2}%
\frac{\delta \langle a^{\dag }a^{\dag }( T)\rangle }{\delta \mathcal{Q}( s) }%
+\frac{\delta \langle a^{\dag }a( T) \rangle }{\delta \mathcal{Q}( s)}+\frac{%
1}{2}\frac{\delta \langle aa(T) \rangle }{\delta \mathcal{Q}( s)}, \\
&\frac{\delta \mathbf{V}_{12}( T) }{\delta \mathcal{Q}( s)} =\frac{i}{2}%
\frac{\delta \langle a^{\dag }a^{\dag }( T)\rangle }{\delta \mathcal{Q}( s) }%
-\frac{i}{2}\frac{\delta \langle aa( T) \rangle }{\delta \mathcal{Q}( s) }, \\
&\frac{\delta \mathbf{V}_{13}( T) }{\delta \mathcal{Q}( s)} =\frac{1}{2}%
\frac{\delta \langle a^{\dag }b^{\dag }( T)\rangle }{\delta \mathcal{Q}( s) }%
+\frac{1}{2}\frac{\delta \langle a^{\dag }b( T) \rangle }{\delta \mathcal{Q}%
( s)}+\frac{1}{2}\frac{\delta \langle ab^{\dag }(T)\rangle }{\delta \mathcal{%
Q}( s) }+\frac{1}{2}\frac{\delta \langle ab( T) \rangle }{\delta \mathcal{Q}%
( s)}, \\
&\frac{\delta \mathbf{V}_{14}( T) }{\delta \mathcal{Q}( s)} =\frac{i}{2}%
\frac{\delta \langle a^{\dag }b^{\dag }( T)\rangle }{\delta \mathcal{Q}( s)}-%
\frac{i}{2}\frac{\delta\langle a^{\dag }b( T) \rangle }{\delta \mathcal{Q}(
s)}+\frac{i}{2}\frac{\delta\langle ab^{\dag }(T) \rangle }{\delta \mathcal{Q}%
( s) }-\frac{i}{2}\frac{\delta\langle ab( T)\rangle }{\delta \mathcal{Q}( s)},
\\
&\frac{\delta \mathbf{V}_{22}( T) }{\delta \mathcal{Q}( s)} =-\frac{1}{2}%
\frac{\delta \langle a^{\dag }a^{\dag }( T)\rangle }{\delta \mathcal{Q}( s) }%
+\frac{\delta \langle a^{\dag }a( T) \rangle }{\delta \mathcal{Q}( s)}-\frac{%
1}{2}\frac{\delta \langle aa( T) \rangle }{\delta \mathcal{Q}( s)}, \\
&\frac{\delta \mathbf{V}_{23}( T) }{\delta \mathcal{Q}( s)} =\frac{i}{2}%
\frac{\delta \langle a^{\dag }b^{\dag }( T)\rangle }{\delta \mathcal{Q}( s)}+%
\frac{i}{2}\frac{\delta\langle a^{\dag }b( T) \rangle }{\delta \mathcal{Q}(
s) }-\frac{i}{2}\frac{\delta \langle ab^{\dag }(T) \rangle }{\delta \mathcal{%
Q}( s) }-\frac{i}{2}\frac{\delta \langle ab( T) \rangle }{\delta \mathcal{Q}%
( s) }, \\
&\frac{\delta \mathbf{V}_{24}( T) }{\delta\mathcal{Q}( s)} =-\frac{1}{2}%
\frac{\delta \langle a^{\dag }b^{\dag }( T)\rangle }{\delta \mathcal{Q}( s)}+%
\frac{1}{2}\frac{\delta\langle a^{\dag }b( T) \rangle }{\delta \mathcal{Q}(
s)}+\frac{1}{2}\frac{\delta \langle ab^{\dag }(T) \rangle }{\delta\mathcal{Q}%
( s)}-\frac{1}{2}\frac{\delta \langle ab( T) \rangle }{\delta \mathcal{Q}( s)%
}, \\
&\frac{\delta \mathbf{V}_{33}( T) }{\delta \mathcal{Q}( s)} =\frac{1}{2}%
\frac{\delta \langle b^{\dag }b^{\dag }( T)\rangle }{\delta \mathcal{Q}( s)}+%
\frac{\delta \langle b^{\dag }b( T) \rangle }{\delta \mathcal{Q}( s)}+\frac{1%
}{2}\frac{\delta\langle bb( T) \rangle }{\delta \mathcal{Q}( s)}, \\
&\frac{\delta \mathbf{V}_{34}( T) }{\delta \mathcal{Q}( s)} =\frac{i}{2}%
\frac{\delta\langle b^{\dag }b^{\dag }( T) \rangle }{\delta \mathcal{Q}( s) }%
-\frac{i}{2}\frac{\delta\langle bb( T) \rangle }{\delta \mathcal{Q}( s)}, \\
&\frac{\delta \mathbf{V}_{44}( T) }{\delta \mathcal{Q}( s)} =-\frac{1}{2}%
\frac{\delta\langle b^{\dag }b^{\dag }( T)\rangle }{\delta \mathcal{Q}( s) }+%
\frac{\delta \langle b^{\dag }b( T) \rangle }{\delta \mathcal{Q}( s)}-\frac{1%
}{2}\frac{\delta \langle bb( T)\rangle }{\delta \mathcal{Q}( s)}.
\end{align}
\end{subequations}
The variation for other covariance-matrix elements can be obtained based on the Hermitian conjugate relations.

It can be seen from Eq.~(\ref{A3}) that, to obtain the result in Eq.~(\ref{subeq2}), we need to further calculate the variations of these second-order moments with respect to $\Omega(s)$ and $\phi(s)$. This can be achieved by taking the variation with respect to $\Omega(s)$
and $\phi(s)$ on both sides of Eq.~(\ref{eq5.2}), namely,
\begin{equation}\label{eq121}
\frac{\delta \dot{\mathbf{X}}( t) }{\delta \mathcal{Q} ( s) } =\mathbf{M}( t)
\frac{\delta \mathbf{X}( t) }{\delta \mathcal{Q}( s) }+\frac{\delta \mathbf{M%
}( t) }{\delta \mathcal{Q}( s) }\mathbf{X}( t) +\frac{\delta \mathbf{N}( t)
}{\delta \mathcal{Q}( s) }.
\end{equation}
Since $\mathcal{Q}( s)$ is a function of $s$, $\delta \dot{\mathbf{X}}( t) /\delta \mathcal{Q} ( s) $ in Eq.~(\ref{eq121}) can be expressed as $\frac{d}{dt}\frac{\delta {\mathbf{X}}( t) }{\delta \mathcal{Q} ( s) }$~\cite{Greiner1996FQ}. Under the initial
condition $\delta \mathbf{X}( 0)/\delta \mathcal{Q}( s)=[0,...,0]_{10\times1}^{\text{T}}$, the solution of  Eq.~(\ref{eq121}) at the target time $T$ can be obtained as
\begin{align}\label{eq8}
\frac{\delta \mathbf{X}(T)}{\delta \mathcal{Q}(s)} &=\mathbf{U}(T)\frac{%
\delta \mathbf{X}(0)}{\delta \mathcal{Q}(s)}+\mathbf{U}(T)\int_{0}^{T}d\tau
\mathbf{U}^{-1}(\tau )\left[ \frac{\delta \mathbf{M}(\tau )}{\delta \mathcal{%
Q}(s)}\mathbf{X}(\tau )+\frac{\delta \mathbf{N}(\tau )}{\delta \mathcal{Q}(s)%
}\right]   \notag \\
&=\mathbf{U}(T)\int_{0}^{T}d\tau \mathbf{U}^{-1}(\tau )\left[ \frac{\delta
\mathbf{M}(\tau )}{\delta \mathcal{Q}(\tau )}\frac{\delta \mathcal{Q}(\tau )%
}{\delta \mathcal{Q}(s)}\mathbf{X}(\tau )+\frac{\delta \mathbf{N}(\tau )}{%
\delta \mathcal{Q}(\tau )}\frac{\delta \mathcal{Q}(\tau )}{\delta \mathcal{Q}%
(s)}\right]   \notag \\
&=\mathbf{U}(T)\mathbf{U}^{-1}(s)\left[ \frac{\delta \mathbf{M}(s)}{\delta
\mathcal{Q}(s)}\mathbf{X}(s)+\frac{\delta \mathbf{N}(s)}{\delta \mathcal{Q}%
(s)}\right] ,
\end{align}
where $\mathbf{U}(t)$ satisfies $\dot{\mathbf{U}}(t)=\mathbf{M}(t)\mathbf{U}%
(t) $ with the initial value $\mathbf{U}(0)=I$.

To know the explicit expressions for $\delta \mathbf{M}(s)/\delta \mathcal{Q}( s)$ and $%
\delta \mathbf{N}(s)/\delta \mathcal{Q}( s)$ in Eq.~(\ref{eq8}), we need to obtain the values of $%
\delta \alpha(s)/\delta \mathcal{Q}( s)$, $\delta \beta(s)/\delta
\mathcal{Q}( s)$, and their Hermitian conjugate.  Taking the variation with respect to $\mathcal{Q}(s)$ on both sides of Eqs.~(\ref{eq2a}) and (\ref{eq2b}) as well as their Hermitian conjugate equations, we have
\begin{equation}
\dot{\mathbf{A}}(t)=\mathbf{W}(t) \mathbf{A}(t)+\mathbf{Q}(t),
\label{B2}
\end{equation}%
where
\begin{equation}
\mathbf{A}(t)=\left(\frac{\delta \alpha (t)}{\delta \mathcal{Q}(s)},\frac{\delta \beta (t)}{\delta \mathcal{Q}(s)},\frac{\delta \alpha
^{\ast }(t)}{\delta \mathcal{Q}(s)},\frac{\delta \beta ^{\ast }(t)}{\delta \mathcal{Q}(s)}\right)^{\text{T}},
\end{equation}%
\begin{equation}
\mathbf{W}(t)=\left(
\begin{array}{cccc}
-i\Delta (t)-\frac{\kappa }{2} & 0 & -ig_{0}\alpha & -ig_{0}\alpha \\
-ig_{0}\alpha ^{\ast } & -ig_{0}\alpha & -i\omega _{\mathrm{m}}-\frac{\gamma}{2} & 0 \\
0 & i\Delta (t)-\frac{\kappa }{2} & ig_{0}\alpha ^{\ast } & ig_{0}\alpha ^{\ast } \\
ig_{0}\alpha ^{\ast } & ig_{0}\alpha & 0 & i\omega _{\mathrm{m}}-\frac{\gamma}{2}%
\end{array}%
\right) ,
\end{equation}%
and
\begin{equation}
\mathbf{Q}(t)=\left(
\begin{array}{c}
i\frac{\delta \Omega (t)}{\delta \mathcal{Q}(s)}e^{-i\phi (t)}+i\Omega (t)\frac{\delta \lbrack
e^{-i\phi (t)}]}{\delta \mathcal{Q}(s)} \\
0 \\
-i\frac{\delta \Omega (t)}{\delta \mathcal{Q}(s)}e^{i\phi (t)}-i\Omega (t)\frac{\delta \lbrack
e^{i\phi (t)}]}{\delta \mathcal{Q}(s)} \\
0%
\end{array}%
\right) .
\end{equation}%
The solution to Eq.~(\ref{B2}) can be expressed as
\begin{equation}
\mathbf{A}(t)=\Lambda (t) \mathbf{A}(0)+\Lambda
(t)\int_{0}^{t}\Lambda ^{-1}(\tau )\mathbf{Q}(\tau )d\tau ,  \label{B6}
\end{equation}%
where $ \mathbf{A}(0)=[0,0,0,0]^{\text{T}}$ and $\Lambda (t)$ satisfies the equation $%
\dot{\Lambda}(t)=\mathbf{W}(t)\Lambda (t)$ with the initial value $\Lambda (0)=I$. Then, we have
\begin{subequations}
\begin{align}
&\left(\frac{\delta \alpha (s)}{\delta \Omega(s)},\frac{\delta \beta (s)}{\delta \Omega(s)},\frac{\delta \alpha
^{\ast }(s)}{\delta \Omega(s)},\frac{\delta \beta ^{\ast }(s)}{\delta \Omega(s)}\right)^{\text{T}}=\left(\frac{1}{2}ie^{-i\phi
(s)},0,-\frac{1}{2}ie^{i\phi (s)},0\right)^{\text{T}}, \\
&\left(\frac{\delta \alpha (s)}{\delta \phi(s)},\frac{\delta \beta (s)}{\delta \phi(s)},\frac{\delta \alpha
^{\ast }(s)}{\delta \phi(s)},\frac{\delta \beta ^{\ast }(s)}{\delta \phi(s)}\right)^{\text{T}}=\left(\frac{1}{2}\Omega
(s)e^{-i\phi (s)},0,\frac{1}{2}\Omega (s)e^{i\phi (s)},0\right)^{\text{T}}.
\end{align}
\end{subequations}
Thus, we obtain the variation of the displacement amplitudes $\alpha(s)$ and $\beta(s)$ with respect to the driving
amplitude $\Omega(s)$ and phase $\phi(s)$, respectively. Based on the value of $\delta [\Delta X_{b}^{2}( \theta,T)]
/ \delta\mathcal{Q} ( s) $, we can further perform the gradient-descent algorithm until a satisfactory value of the loss function $\Delta X_{b}^{2}( \theta,T)$ is obtained.
\end{widetext}


\begin{thebibliography}{99}
\bibitem{check} We checked the correctness of the pulsed fields
by back-substituting the pulsed field data obtained after the last iteration into the system evolution. We verified the results by either solving the covariance matrix $\mathbf{V}$ or solving the quantum master equation~(%
\ref{Eq45}) under the replacement of $H_{\mathrm{dis}}(t) \rightarrow H_{%
\mathrm{lin}}(t)$. We find that the dynamic evolution of the squeezing degree $\mathcal{%
S}_{b}$ is consistent with the results obtained in this paper.
\end{thebibliography}
\end{document}